\documentclass[aps,floatfix,showpacs]{revtex4}
\usepackage{graphicx}
\usepackage{amssymb}
\usepackage{amsmath}
\usepackage[english,activeacute] {babel}
\def\mapright#1#2{\smash{
     \mathop{-\!\!\!-\!\!\!-\!\!\!\rightarrow}\limits^{#1}_{#2}}}

\begin{document}

\title{Flavor changing neutral currents in 331 models}

\author{J.M.\ Cabarcas$^a$, D.\ G\'omez Dumm$^a$ and R.\ Martinez$^b$}

\affiliation{ $^a$ IFLP, CONICET -- Dpto.\ de F\'{\i}sica, Universidad
Nacional de La Plata, \\ C.C. 67, 1900 La Plata, Argentina. \\
$^b$ Dpto.\ de F\'{\i}sica, Universidad Nacional, Bogot\'a, Colombia.}

\begin{abstract}
We carry out a general analysis of tree level flavor changing neutral
currents in the context of 331 models, considering arbitrary quark and
gauge boson mixing matrices. The results are applied to definite textures
of quark mass matrices, and differences between various 331 scenarios are
pointed out.
\end{abstract}

\maketitle

\section{Introduction}

It is well known that in the Standard Model (SM) flavor changing neutral
currents (FCNC) are forbidden at tree level, thus FCNC-mediated processes
are only allowed through one or more loop corrections. On the experimental
side, even if the sizes of FCNC in the $u-c$, $b-s$, $s-d$ and $b-d$ sectors
are found to be in good agreement with SM predictions, there is still room
available for the presence of new physics.

In this paper we address the problem of FCNC in the context of the so-called
331 models, in which the SM gauge group is enlarged to $SU(3)_C\otimes
SU(3)_L \otimes U(1)_X$~\cite{Pisano:1991ee}. This extension of the gauge
group implies the presence of a new neutral gauge boson $Z'$, which in
general gives rise to tree-level FCNC. The existence of $Z'$ gauge bosons
occurs not only in 331 models but in many popular extensions of the SM, and
in general $Z'$ searches are part of the main experimental programs in high
energy physics. Among the various extensions of the SM, 331 models have the
attractive feature of addressing the problem of fermion family replications.
The puzzle is partially solved through the requirement of anomaly
cancellations~\cite{Pisano:1991ee,Ng:1992st,kiyan:2002qw,Diaz:2004fs}, since
in the 331 framework the number of quarks families turns out to be related
to the number of colors. In addition, 331 models show a rich phenomenology,
which includes the presence of new scalars and heavy quarks, and offers the
possibility of see-saw mechanisms to generate neutrino
masses~\cite{Montero:2000rh}, dark matter
candidates~\cite{Fregolente:2002nx}, gauge coupling
unification~\cite{Diaz:2005bw}, etc.

It is important to emphasize that in 331 models it is not possible to fit
all SM quark and lepton families in multiplets having the same quantum
numbers. As a consequence, the corresponding $Z'$ couplings are not
universal for all fermion flavors, which leads to tree-level
FCNC~\cite{Liu:1993gy,GomezDumm:1994tz,Rodriguez:2004mw,Promberger:2007py,
Cabarcas:2007my,Promberger:2008xg,Martinez:2008jj}. Moreover, the
couplings of left-handed fermions to the ``standard'' $Z$-like gauge boson
also include tree-level FCNC, both through $Z-Z'$ mixing and through the
mixing between ordinary and exotic quarks (the latter, in 331 models where
exotic quarks have ordinary electric charges). The sizes of the couplings
depend on the angles and phases of the (left-handed) up and down quark
mixing matrices $V_L^u$ and $V_L^d$, which become separately observable.
It is noticeable that, because of the unitarity of these matrices,
predictions for FCNC observables in 331 models are in general related to
each other~\cite{Promberger:2007py,Cabarcas:2007my,Cabarcas:2008ys}.
Moreover, the extensions of the SM fermion and scalar sectors lead to new
CP violation sources when quark and/or lepton mixing are taken into
account. Experimentally, the nonstandard neutral currents could be
identified e.g.\ at the LHC by looking at the process $pp\to Z'\to e^+
e^-$: performing specific kinematic cuts on the outgoing electrons, it
would be possible to reduce background so as to distinguish the $Z'$
current within 331 models from other theories that include physics beyond
the SM~\cite{Dittmar:2003ir}.

In this work we present a general analysis of the various sources of tree
level FCNC in 331 models, showing the relations between the involved quark
mixing matrix elements. In particular, we consider the case in which
nonstandard quarks carry ordinary $2/3$ and $-1/3$ electric charges, thus
they mix with ordinary fermions and lead to tree level FCNC both through
$Z$- and $Z'$-quark couplings. In this case one also finds further neutral
heavy gauge bosons, which mix with the $Z$ and $Z'$ states and may provide a
source of CP violation. The aim of this paper is to present an overall
analysis of FCNC sources, generalizing previous results obtained in
particular 331 theories. Our results are compared with model-independent
studies that do not include $Z$-mediated
FCNC~\cite{Nardi:1992nq,Langacker:2000ju}, and with previous analyses within
331 models in which exotic quarks have nonstandard electric charges, and
therefore do not mix with ordinary
fermions~\cite{Liu:1993gy,Promberger:2007py,Perez:2004jc}. The mixing
between CP-even and CP-odd neutral gauge bosons is also taken into account.

The article is organized as follows: in Sect.~II we introduce the models
under study. In Sect.~III we analyze the couplings between fermions and
gauge bosons, focusing on the study of FCNC. The case in which fermion
mixing matrices show family hierarchies is considered. An application of
this formalism to analyze different contributions to neutral meson mass
differences is presented in Sect.~IV. Finally, Sect.~V includes a brief
summary of our results.

\section{331 Models}

As stated above, in 331 models the SM gauge group is enlarged to
$SU(3)_C\otimes SU(3)_L\otimes U(1)_X$. Left-handed fermions are organized
into $SU(3)_L$ triplets, which in general include the standard quarks and
leptons, as well as new exotic particles. The criterion of anomaly
cancellation leads to some constraints in the allowed fermion
representations ({\bf 3} or $\bar {\bf 3}$) and the fermion quantum
numbers.

To describe the particle content of the model we start by defining the
electric charge operator, which can be written as a linear combination of
the diagonal generators of the group:
\begin{equation}
Q\ = \ T_3\,+\,\beta\, T_8\,+\,X\ .
\end{equation}
Here $\beta$ is a parameter that characterizes the specific particle
structure and quantum numbers of the model. For arbitrary $\beta$ the quark
content of the multiplets consistent with anomaly cancellation requirements
and a minimal number of exotic particles~\cite{Diaz:2004fs} is the
following:
\begin{eqnarray}
& & {\cal Q}_{iL} =
\left( \begin{array}{ccc}
d_{i} & -\, u_{i} & J_{i}
\end{array} \right)^T_L \ \sim \
\left(\bar {\bf 3}\; , \;\frac{1}{6}+\frac{\beta}{2\sqrt{3}}\right)\, , \ \ i=1,2
\nonumber \\
& & {\cal Q}_{3L} =
\left( \begin{array}{ccc}
d_{3} & u_{3} & J_{3}
\end{array} \right)^T_L \ \sim \
\left({\bf 3}\; , \;\frac{1}{6}-\frac{\beta}{2\sqrt{3}}\right)
\nonumber \\
& & u_{jR} \ \sim \ \left({\bf 1}\; , \;\frac{2}{3}\right)\, ,
\ \ d_{jR} \ \sim \ \left({\bf 1}\; , \;-\frac{1}{3}\right)\, , \ \ j=1,2,3
\nonumber \\
& & J_{iR} \ \sim \ \left({\bf 1}\; , \;\frac{1}{6} + \frac{\sqrt{3}\beta}{2}\right)\,
, \ \ i=1,2\ , \qquad J_{3R} \ \sim \ \left({\bf 1}\; ,
\;\frac{1}{6}-\frac{\sqrt{3}\beta}{2}\right)\ ,
\label{content}
\end{eqnarray}
where entries in the parentheses label the $SU(3)_L$ representation and the
$X$ quantum number. It is easy to see that the electric charges of exotic
quarks $J$ are given by $Q_{J_1}=Q_{J_2}=1/6+\sqrt{3}\beta/2$ and
$Q_{J_3}=1/6-\sqrt{3}\beta/2$. Therefore, the sole values of $\beta$ that
lead to standard electric charges for these fermions are $\beta=\pm
1/\sqrt{3}$~\cite{Foot:1994ym,Ozer:1995xi}. Anomaly cancellation requires
that two of the quark families lie in the ${\bf 3}$, while the remaining
family, together with the leptons, lies in the $\bar {\bf 3}$ (or
viceversa). Left-handed leptons are accommodated into an $SU(3)_L$ triplet,
in the same way as the quarks ${\cal Q}_{3L}$, and this can be done with or
without the inclusion of nonstandard
particles~\cite{Liu:1993gy,Foot:1992rh,Montero:2001tq}. To get an
anomaly-free theory it is necessary to have the same number of fermions in
the ${\bf 3}$ and the $\bar {\bf 3}$. Therefore, if the model has a quark
content as shown in Eq.~(\ref{content}), the number of lepton families has
to be the same as the number of quark colors. In what follows we will not
discuss the features of lepton interactions, focusing on the analysis of
flavor changing currents in the quark sector.

The gauge bosons associated with $SU(3)_L$ transform according to the
adjoint representation of the group, and can be written as
\begin{equation}
\mathbf{W}_\mu \ = \ W_\mu ^a \frac{\lambda^a}{2} \ = \ \frac 12\left(
\begin{array}{ccc}
W_\mu^3 + \frac 1{\sqrt{3}}W_\mu ^8 & \sqrt{2}\, W_\mu ^{+} &
\sqrt{2}\, K_{1\mu} \\
\sqrt{2}\, W_\mu ^{-} & -W_\mu ^3 + \frac 1{\sqrt{3}} W_\mu ^8 &
\sqrt{2}\, \bar K_{2\mu} \\
\sqrt{2}\, \bar K_{1\mu} & \sqrt{2}\, K_{2\mu} & -\frac 2
{\sqrt{3}}W_\mu ^8
\end{array}
\right) \ ,
\label{3}
\end{equation}
where $\lambda^a$ are the Gell-Mann matrices, and the electric charges of
$K_1$ and $K_2$ are given by $Q_1 = 1/2+(\sqrt{3}\beta)/2$ and $Q_2 =
1/2-(\sqrt{3}\beta)/2$, respectively. One has in addition a gauge field
$B_\mu$ associated with $U(1)_X$; this boson is a singlet under $SU(3)_L$
and has zero electric charge.

In general, it is convenient to rotate the neutral gauge bosons $W^3$,
$W^8$ and $B$ into new states $A$, $Z$ and $Z'$, given by
\begin{eqnarray}
\left(\begin{array}{c}
A \\ Z \\ Z^\prime
\end{array}\right) =
\left( \begin{array}{ccc}
S_W & \beta S_W & C_W\sqrt{1-\beta^2 T_W^2} \\
C_W & -\,\beta S_W T_W & -\,S_W\sqrt{1-\beta^2 T_W^2} \\
0 & -\sqrt{1-\beta^2 T_W^2} & \beta T_W
\end{array} \right)
\left( \begin{array}{c}
W^3 \\ W^8 \\ B
\end{array} \right) \ ,
\label{bosgauneu}
\end{eqnarray}
where the ``Weinberg angle'' $\theta_W$ is defined by $T_W = \tan\theta_W
= g'/\sqrt{g^2+\beta^2{g'}^2}$, $g$, $g'$ being the coupling constants
associated to the groups $SU(3)_L$ and $U(1)_X$ respectively ($S_W =
\sin\theta_W$, etc.). In the new basis, $A_\mu$ (the photon) is the gauge
boson corresponding to the generator $Q$, while $Z_\mu$ can be identified
with the SM $Z$ boson. As in the SM, the extended electroweak symmetry is
spontaneously broken in 331 models by the presence of elementary scalars
having nonzero vacuum expectation values~\cite{Nguyen:1998ui,Tully:1998wa,
Diaz:2003dk,Ponce:2002sg,Dong:2006mg}. The symmetry breakdown follows a
hierarchy
\begin{equation}
SU(3)_L\otimes U(1)_X \ \mapright{\hspace{-.9cm}
V\hspace{-.9cm}}{}\ SU(2)_L\otimes U(1)_Y \
\mapright{\hspace{-.9cm} v
\hspace{-.9cm}}{}\ U(1)_Q \ \ ,
\label{jerarq}
\end{equation}
in which two VEV scales $V$ and $v$, with $V\gg v$, are introduced. The
photon is kept as the only massless gauge boson, while the remaining neutral
gauge bosons get mixed. In this way, $Z$ and $Z'$ turn out to be only
approximate mass eigenstates.

In the following we will mainly focus on the models with $\beta =
\pm1/\sqrt{3}$, in which there are no quarks with exotic charges and the
possible existence of fermion and gauge boson mixing and FCNC is maximized.
These theories include the so-called
``minimal''~\cite{Foot:1994ym,Diaz:2003dk} and
``economical''~\cite{Ponce:2002sg,Dong:2006mg} 331 models, for which
different phenomenological analyses have been carried out in the last few
years. The case of 331 models with $\beta=\pm\sqrt{3}$~\cite{Pisano:1991ee}
is briefly addressed in Sect.~IV.

\section{Fermion couplings to neutral gauge bosons.}

We are interested in the analysis of neutral currents driven by gauge bosons
in the framework of 331 models with $\beta = \pm1/\sqrt{3}$. Taking into
account the quark content in Eq.~(\ref{content}), it is easy to obtain the
corresponding interaction Lagrangian for the $Z$ and $Z'$ gauge bosons. One
gets
\begin{eqnarray}
{\cal L}_{\rm NC}^{(Z,Z')} & = & -\frac{g}{2C_W}
\left\{\bar U^0 \gamma_\mu\left[ \left(C_W^2-\frac{S_W^2}{3}\right)P_L
-\frac{4S_W^2}{3}P_R\right]U^0 -
\sum_{i=1}^{n_T}\frac{4S_W^2}{3} \bar{T}_{i}^0\gamma_\mu T_{i}^0 \right.
\nonumber\\
& & \left. +\ \bar D^0 \gamma_\mu\left[
\left(-C_W^2-\frac{S_W^2}{3}\right)P_L + \frac{2S_W^2}{3}P_R\right] D^0 +
\sum_{i=1}^{n_B}\frac{2S_W^2}{3} \bar{B}_{i}^0\gamma_\mu B_{i}^0\right\}
Z^{\mu}
\nonumber\\
& & - \ \frac{g}{2 C_W}\; \frac{1}{\sqrt{3 C_W^2 - S_W^2}}
\left\{\bar U^0\gamma_\mu \left[\left(\Lambda C_W^2 \pm
\frac{S_W^2}{3} \right) P_L \pm \frac{4S_W^2}{3}P_R \right] U^0
+ \right. \nonumber \\
& & \pm \sum_{i=1}^{n_T} \bar{T}_{i}^0 \gamma_\mu \left[\left(-2
C_W^2+\frac{4S_W^2}{3}\right)
P_L + \frac{4S_W^2}{3}P_R\right]T_{i}^0 +
\bar D^0\gamma_\mu \left[\left(C_W^2\Lambda \pm
\frac{S_W^2}{3}\right) P_L \mp \frac{2S_W^2}{3} P_R \right] D^0
\nonumber \\
& & \left. \pm
\sum_{i=1}^{n_B}\bar{B}_{i}^0 \gamma_\mu \left[\left(2C_W^2 -
\frac{2S_W^2}{3}\right) P_L -
\frac{2S_W^2}{3}P_R\right] B_{i}^0\right\} {Z'}^{\mu} \ \ ,
\label{lagquarks}
\end{eqnarray}
where $P_{L,R} = (1\mp\gamma_5)/2$, and we have defined the matrices
$D^0=(d_1^0\ d_2^0\ d_3^0)^T$ , $U^0=(u_1^0\ u_2^0\ u_3^0)^T$ and
$\Lambda={\rm diag}(1,1,-1)$ (flavor space). Superindices $0$ indicate
that quarks are in the weak interaction basis. Quarks $J$ in
Eq.~(\ref{content}) have been renamed by $T_{i}$ and $B_{i}$ to emphasize
that they carry the same electric charges as ordinary top and bottom. The
number of these exotic quarks, $n_T$ and $n_B$, can be 1 or 2 depending on
the choice of the parameter $\beta$: for $\beta = 1/\sqrt{3}$ one has two
up-type quarks $T_{1,2}$ (then $n_T=2$) and one down-type quark $B_{1}$
($n_B=1$), while for $\beta = -1/\sqrt{3}$ one has $n_T=1$, $n_B=2$. Due
to the particular representation structure of the quark sector, the
interactions between the $Z'$ boson and the left-handed quarks include the
matrix $\Lambda$, which is not proportional to the identity. This leads to
tree-level FCNC after rotating the quarks to the mass eigenstate basis. In
addition, notice that there is no universality between the interactions of
ordinary and exotic left-handed quarks, both for $Z$ and $Z'$ currents. In
this way, the mixing between ordinary and exotic quarks leads to further
sources of FCNC.

It is worth to notice that for $\beta = \pm 1/\sqrt{3}$ the gauge bosons
$K_1$ and $K_2$ will have electric charges 0 or 1. Therefore, in general
they will mix with the remaining neutral gauge bosons and the $W^+$. For
$\beta = 1/\sqrt{3}$ one gets a neutral gauge boson $K_2^0$, and the
corresponding weak neutral current reads
\begin{eqnarray}
{\cal L}_{\rm NC}^{(K_2)}\ =\ -\;\frac{g}{\sqrt{2}}\left(\,
\bar B_1^0\gamma_\mu P_L d_3^0 \ + \ \sum_{i=1}^2
\bar u_i^0\gamma_\mu P_L T_i^0 \, \right) K_2^\mu\ + \ {\rm h.c.} \ \ ,
\end{eqnarray}
while for $\beta = - 1/\sqrt{3}$ the neutral state is $K_1$, leading to the
neutral current
\begin{eqnarray}
{\cal L}_{\rm NC}^{(K_1)}\ =\ -\;\frac{g}{\sqrt{2}}\left(\,
\bar u_3^0\gamma_\mu P_L T_1^0 \ - \ \sum_{i=1}^2
\bar B_i^0\gamma_\mu P_L d_i^0 \,
\right) K_1^\mu \ + \ {\rm h.c.} \ \ .
\end{eqnarray}

Let us take for definiteness $\beta = 1/\sqrt{3}$. As one can see from
Eq.~(\ref{3}), the complex fields $K_2$ and $\bar K_2$ are in fact
combinations of the neutral gauge bosons $W^6$ and $W^7$. These are naturally
defined as CP-even and CP-odd states respectively, since these definitions
lead to a CP-invariant weak current in the interaction basis. Thus, once the
photon has been identified, the remaining four neutral gauge boson mass
eigenstates will be written in general as linear combinations of $Z$, $Z'$,
$W^6$ and $W^7$, and the mixing between $W^7$ and any of the other will
imply nonconservation of the CP symmetry in the theory. An equivalent
reasoning applies to $K_1$, $\bar K_1$, $W^4$ and $W^5$ in the case $\beta =
-1/\sqrt{3}$.

Now it is convenient to organize the nonstandard fermions together with the
up- and down-like quarks $U_i^0$ and $D_i^0 $ into four or five-component
fermion vectors ${\cal U}$, ${\cal D}$ defined as
\begin{equation}\label{vectors}
{\cal U}^0 \ = \ \left(
\begin{array}{c}
U^0 \\ T^0
\end{array}
\right) \ , \qquad
{\cal D}^0 \ = \ \left(
\begin{array}{c}
D^0 \\ B^0
\end{array}
\right) \ \ ,
\end{equation}
where $T^0 = (T^0_1 \ T^0_2)^T$, $B^0 = B^0_1$ for $\beta = 1/\sqrt{3}$ and
$T^0 = T^0_1$, $B^0 = (B^0_1 \ B^0_2)^T$ for $\beta = - 1/\sqrt{3}$. With
this notation the neutral currents driven by the $Z$ and $Z'$ bosons can be
written as
\begin{eqnarray}
{\cal L}_{NC}^{(Z,Z')} & = & - \frac{g}{2C_W} \sum_{f = {\cal U},{\cal D}}
\left[ \bar{f}^0 \gamma^\mu (\epsilon_{f_L}^{(1)} P_L +
\epsilon_{f_R}^{(1)}P_R) f^0 \, Z_\mu \ + \
\frac{1}{\sqrt{3 C_W^2 - S_W^2}}\,
\bar{f}^0 \gamma^\mu  (\epsilon_{f_L}^{(2)} P_L +
\epsilon_{f_R}^{(2)}P_R) f^0 \, Z'_\mu
\right] \ \ ,
\label{currzzp}
\end{eqnarray}
where
\begin{eqnarray}
& & \epsilon^{(1)}_{{\cal U}_L}=\left(C_W^2 -
\frac{S_W^2}{3}\right)\; \openone \ - \ \left(\begin{array}{cc}
0_{3\times 3} & \\
& \openone_{n_T\times n_T}
\end{array}\right)
\ , \qquad
\epsilon^{(1)}_{{\cal
D}_L}=\left(- C_W^2 - \frac{4S_W^2}{3}\right)\; \openone
\ + \
\left(\begin{array}{cc}
0_{3\times 3} & \\
& \openone_{n_B\times n_B}
\end{array}\right)
\label{eps1ud} \\
& & \epsilon^{(1)}_{{\cal U}_R}=-\frac{4S_W^2}{3}\; \openone
\ , \qquad
\epsilon^{(1)}_{{\cal D}_R} = \frac{2S_W^2}{3}\; \openone
\\
& & \epsilon^{(2)}_{{\cal
U}_L}=\left(C_W^2 \pm \frac{S_W^2}{3}\right)\; \openone
\ - \ 2\,C_W^2
\left(\begin{array}{cc}
0_{2\times 2} & \\
& \openone_{(n_T+1) \times (n_T+1)}
\end{array}\right)
\ + \ (C_W^2 \mp 2C_W^2 \pm S_W^2)
\left(\begin{array}{cc}
0_{3\times 3} & \\
& \openone_{n_T\times n_T}
\end{array}\right)
\label{eps2u} \\
& & \epsilon^{(2)}_{{\cal
D}_L}=\left(C_W^2 \pm \frac{S_W^2}{3}\right)\; \openone
\ - \ 2\,C_W^2
\left(\begin{array}{cc}
0_{2\times 2} & \\
& \openone_{(n_B+1) \times (n_B+1)}
\end{array}\right)
\ + \ (C_W^2 \pm 2C_W^2 \mp S_W^2)
\left(\begin{array}{cc}
0_{3\times 3} & \\
& \openone_{n_B\times n_B}
\end{array}\right)
\label{eps2d} \\
& & \epsilon^{(2)}_{{\cal U}_R} = \pm\frac{4S_W^2}{3}\; \openone
\ , \qquad
\epsilon^{(2)}_{{\cal D}_R} = \mp \frac{2S_W^2}{3}\; \openone \
\end{eqnarray}
(blank entries in the above matrices are zeros). In the same way the neutral
currents involving $K$ vector bosons for $\beta = 1/\sqrt{3}$ are given by
\begin{eqnarray}
{\cal L}_{NC}^{(K)} &  = & - \frac{g}{\sqrt{2}} \sum_{f = {\cal U},{\cal D}}
\left( \bar{f}^0 \gamma^\mu \epsilon_{f_L}^{(3)} P_L f^0 \, {\rm
Re} K_{2\mu} \ + \, i\, \bar{f}^0 \gamma^\mu \epsilon_{f_L}^{(4)} P_L
f^0 \, {\rm Im} K_{2\mu} \right)
\label{currk}
\end{eqnarray}
with
\begin{eqnarray}
& & \epsilon^{(3)}_{{\cal
U}_L} = \left(\begin{array}{ccc}
0_{2\times 2} & & \openone_{2\times 2}\\
& 0 & \\
\openone_{2\times 2} & & 0_{2\times 2}
\end{array}\right) \ ,
\qquad \epsilon^{(3)}_{{\cal
D}_L} = \left(\begin{array}{ccc}
0_{2\times 2} & & \\
& 0 & 1 \\
& 1 & 0
\end{array}\right) \ , \\
& & \epsilon^{(4)}_{{\cal
U}_L} = \left(\begin{array}{ccc}
0_{2\times 2} & & \openone_{2\times 2}\\
& 0 & \\
-\;\openone_{2\times 2} & & 0_{2\times 2}
\end{array}\right) \ ,
\qquad \epsilon^{(4)}_{{\cal
D}_L} = \left(\begin{array}{ccc}
0_{2\times 2} & & \\
& 0 & -1 \\
& 1 & 0
\end{array}\right) \ ,
\end{eqnarray}
while for $\beta = -1/\sqrt{3}$, replacing $K_2\to K_1$ in
Eq.~(\ref{currk}), one has
\begin{eqnarray}
& & \epsilon^{(3)}_{{\cal
U}_L} = \left(\begin{array}{ccc}
0_{2\times 2} & & \\
& 0 & 1 \\
& 1 & 0
\end{array}\right) \ ,
\qquad \epsilon^{(3)}_{{\cal
D}_L} =  \left(\begin{array}{ccc}
0_{2\times 2} & & -\;\openone_{2\times 2}\\
& 0 & \\
-\;\openone_{2\times 2} & & 0_{2\times 2}
\end{array}\right)\ , \\
& & \epsilon^{(4)}_{{\cal
U}_L} = \left(\begin{array}{ccc}
0_{2\times 2} & & \\
& 0 & 1 \\
& -1 & 0
\end{array}\right) \ ,
\qquad \epsilon^{(4)}_{{\cal
D}_L} =  \left(\begin{array}{ccc}
0_{2\times 2} & & \openone_{2\times 2}\\
& 0 & \\
-\;\openone_{2\times 2} & & 0_{2\times 2}
\end{array}\right)\ .
\end{eqnarray}

Up to now the currents have been written in the interaction basis. In order
to move to the fermion mass eigenstate basis, we introduce unitary matrices
$V^f_{L,R}$, with $f={\cal U,D}$, that diagonalize the fermion mass matrices
arising from the Yukawa couplings in the 331
model~\cite{Diaz:2003dk,Dong:2006mg,Ponce:2002sg,Cotaescu:2008ya}. One has
thus
\begin{equation}
{\cal U}^0_{L,R}\ =\ V^{\cal U}_{L,R}\, {\cal U}_{L,R} \ , \qquad
{\cal D}^0_{L,R}\ =\ V^{\cal D}_{L,R}\, {\cal D}_{L,R} \ ,
\end{equation}
where ${\cal U}$, ${\cal D}$ are quark mass eigenstate vectors. The
matrices $V^f_{L,R}$ will have dimensions $4\times 4$ or $5\times 5$ for
the up or down sector, depending on the choice of the parameter $\beta$.
Given the structure of the couplings, it is useful to write the matrices
$V_L^{\cal U,D}$ in terms of submatrices. For $V_L^{\cal U}$ we define
\begin{equation}
V_L^{\cal U} \ = \ \left(
\begin{array}{lll}
{V_{qq}^{\cal U}}_{(2\times 2)} & {V_{qQ}^{\cal U}}_{(2\times 1)} &
{V_{qJ}^{\cal U}}_{(2\times n_T)} \\
{V_{Qq}^{\cal U}}_{(1\times 2)} & V_{QQ}^{\cal U} &
{V_{QJ}^{\cal U}}_{(1\times n_{T})} \\
{V_{Jq}^{\cal U}}_{(n_{T}\times 2)} &
{V_{JQ}^{\cal U}}_{(n_{T}\times 1)} &
{V_{JJ}^{\cal U}}_{(n_{T}\times n_{T})}
\end{array}
\right) \ ,
\label{submat}
\end{equation}
while same definitions apply to $V_L^{\cal D}$, with the replacements ${\cal
U}\to {\cal D}$, $n_T\to n_B$. The textures of these matrices arise from the
scalar VEVs and the parameters entering the Yukawa sector of the model. Even
if these are in principle unknown, the common belief is that there is a
hierarchy in the masses and mixing angles between the quark families. Thus
for the moment we will take $V^f_{L,R}$ as unknown, and then we will
consider the presence of hierarchies between the different submatrices in
$V_L^{\cal U,D}$. Since right-handed currents are flavor diagonal (as occurs
in the SM), the mixing matrices $V_R^{{\cal U},{\cal D}}$ are not observable
in 331 models. On the contrary, the parameters in $V_L^{\cal U}$ and
$V_L^{\cal D}$ become separately observable due to the non-universality of
the couplings in Eqs.~(\ref{currzzp}) and (\ref{currk}).

The charged currents involving the $W^\pm$ gauge bosons are similar to
those in the SM, i.e.
\begin{equation}
{\cal L}_{\rm CC}^{W}\ =\ -\;\frac{g}{\sqrt{2}}\; \bar U^0\gamma^\mu P_L
D^0\; W_\mu^+ \ + \ {\rm h.c.} \ = \ -\;\frac{g}{\sqrt{2}}\; \bar {\cal
U}^0\gamma^\mu P_L\, \epsilon_W\, {\cal D}^0\; W_\mu^+ \ + \ {\rm h.c.}
\end{equation}
where
\begin{equation}
\epsilon_W \ = \ \left(\begin{array}{cc}
\openone_{3\times 3} & \\
& 0_{n_T\times n_B}
\end{array}\right) \ .
\end{equation}
After rotating to the fermion mass eigenstate basis, in principle all quark
states interact with the $W^\pm$ bosons. We consider in particular the
charged currents involving just the ordinary quarks, which can be written as
\begin{equation}
{\cal L}_{\rm CC}^{(W,\,{\rm ord})}\ =\ -\;\frac{g}{\sqrt{2}}\;
\bar U\gamma^\mu P_L V_{CKM} D\; W_\mu^+\ + \ {\rm h.c.} \ ,
\label{currw}
\end{equation}
where $V_{CKM}$ is the Cabibbo-Kobayashi-Maskawa matrix. In terms of the
above defined submatrices, the latter is given by
\begin{equation}
V_{CKM} \ = \ V_0^{{\cal U}\,\dagger}\; V_0^{\cal D} \ ,
\label{vckm}
\end{equation}
with
\begin{equation}
V_0^f \ = \ \left(
\begin{array}{ll}
{V_{qq}^f}_{(2\times 2)} & {V_{qQ}^{f}}_{(2\times 1)} \\
{V_{Qq}^{f}}_{(1\times 2)} & V_{QQ}^{f}
\end{array}
\right) \ .
\label{vcero}
\end{equation}
Since $V_0^{\cal U,D}$ are $3\times 3$ submatrices of $V_L^{\cal U,D}$, it
is seen that in 331 models with $\beta = \pm 1/\sqrt{3}$ the CKM matrix is
in general not unitary.

In addition one has to take into account the mixing between gauge bosons. We
will concentrate in the neutral sector, since the analysis of FCNC is the
main purpose of this work. For the moment let us assume that the VEVs of the
scalar states in the model can be taken to be real. This assumption depends
on the structure of the scalar potential, and means that there is no
spontaneous violation of the CP symmetry. If this is the case, the state
$\sqrt{2}\,$Im$K$ decouples, becoming an exact mass eigenstate. However, the
vector bosons $Z$, $Z'$ and $\sqrt{2}\,$Re$K$ in general get mixed
\cite{Diaz:2003dk,Ozer:1995xi,Dong:2006mg,Ponce:2002sg}. One can rotate to
the mass eigenstate basis, say $Z_1$, $Z_2$, $Z_3$ (where $Z_1$ is the
ordinary gauge boson seen in high energy experiments) through an orthogonal
mixing matrix $R$:
\begin{equation}
\left(\begin{array}{c}
Z \\ Z' \\ \sqrt2\,{\rm Re}K \end{array}
\right) \ = \ R\;
\left(\begin{array}{c}
Z_1 \\ Z_2 \\ Z_3 \end{array}
\right) \ .
\end{equation}
Thus in the fermion and gauge boson mass basis the neutral currents will
be given by
\begin{eqnarray}
{\cal L}_{NC} & = & - \sum_{f = {\cal U},{\cal D}}
\bigg[ Q_f \bar{f} \gamma^\mu f \, A_\mu \; + \sum_{j,k=1}^3
g_j \bar{f} \gamma^\mu (E_{f_L}^{(j)} P_L +
E_{f_R}^{(j)}P_R) f \, R_{jk}\, Z_{k\mu} \nonumber \\
& & \ \ \ \ \ \ \ \ \ \ \ \ + \; i\,\frac{g}{2}
\bar{f} \gamma^\mu (E_{f_L}^{(4)} P_L +
E_{f_R}^{(4)}P_R) f \, \sqrt{2}\,{\rm Im}K_\mu \bigg] \ ,
\label{neucurr}
\end{eqnarray}
where $Q_f$ stand for the fermion electric charges, the couplings $g_j$
are defined as
\begin{equation}
g_1 = \frac{g}{2C_W} \ ,\qquad
g_2 = \frac{g'}{2\sqrt{3}S_WC_W} =
\frac{g}{2\sqrt{3}C_W\sqrt{C_W^2 - \beta^2 S_W^2}}\ ,\qquad
g_3 = \frac{g}{2} \ ,
\label{gs}
\end{equation}
and the flavor matrices $E_{f_{L,R}}^{(j)}$ are given by
\begin{equation}
E_{f_L}^{(i)} \ = \ V_L^{f\dagger} \epsilon^{(i)}_{f_L} V_L^f
\ , \qquad \qquad
E_{f_R}^{(i)} \ = \ V_R^{f\dagger} \epsilon^{(i)}_{f_R} V_R^f
\ = \ \epsilon^{(i)}_{f_R} \ .
\end{equation}
According to the hierarchy followed by the gauge symmetry breakdown [see
Eq.~(\ref{jerarq})], it is reasonable to expect that $Z_2$, $Z_3$, $T$ and
$B$ get masses of order $V$. These states should decouple for large $V$,
hence the mixing angles between $Z$ and $Z'$, Re$K$ and between ordinary and
exotic quarks are expected to be small. Explicit calculations for definite
scalar potentials and Yukawa couplings can be found in
Refs.~\cite{Diaz:2004fs,Dong:2006mg,Ponce:2002sg}. The mixing angles between
light and heavy neutral gauge bosons are typically suppressed by the ratio
$(v/V)^2\approx (m_{Z_1}/m_{Z_{2,3}})^2$~\cite{Diaz:2004fs,Dong:2006mg},
hence one expects mixing angles of order $\lesssim 0.01$ for masses of
exotic neutral gauge bosons in the TeV range.

Let us concentrate in the FCNC currents that involve the ordinary quarks
$u,c,t,d,s,b$. It is natural to expect that the presence of these FCNC at
the tree level will lead to stringent bounds on the model parameters, and at
the same time to the most promising observable effects of new physics at the
presently available energy scales. Typically the most stringent bounds for
FCNC arise from mass splittings in the neutral meson sectors, therefore we
consider the application of our formalism to these observables in Sect.~IV.
Other flavor-violating processes that have been experimentally
measured/constrained with relatively high precision are e.g.\ $Z$ boson
decays into lepton pairs, quark and lepton radiative decays, and leptonic
and semileptonic decays of neutral mesons.

In our framework, tree level FCNC involving ordinary quarks will arise from
the upper left $3\times 3$ submatrices of the left-handed flavor matrices
$E_{fL}^{(i)}$, which are in general not diagonal. For $i=1,2$ we can also
separate a diagonal part, given by the first terms of $\epsilon^{(i)}_{f_L}$
in Eqs.~(\ref{eps1ud}), (\ref{eps2u}) and (\ref{eps2d}), which remain
unchanged under the fermion state rotation. Thus we write
\begin{equation}
E_{f_L}^{(i)} \ = \ E_{f_L}^{(i,{\rm
diag})} + \Delta E_{f_L}^{(i)} \ .
\label{efl}
\end{equation}
With the definitions introduced in Eqs.~(\ref{submat}) and (\ref{vcero}),
the upper left $3\times 3$ submatrices of $\Delta E_{f_L}^{(i)}$ are given
by
\begin{eqnarray}
\Delta E_{{\cal U}_L, 3\times 3}^{(1)} & = & - \ \left(\begin{array}{cc}
V_{Jq}^{{\cal U}\,\dagger} V_{Jq}^{\cal U} &
V_{Jq}^{{\cal U}\,\dagger} V_{JQ}^{\cal U} \\
V_{JQ}^{{\cal U}\,\dagger} V_{Jq}^{\cal U} \rule{0cm}{.5cm} &
V_{JQ}^{{\cal U}\,\dagger} V_{JQ}^{\cal U}
\end{array}\right) \ = \
- \left(\openone_{3\times 3} \, -
\, V_0^{{\cal U}\,\dagger}\, V_0^{\cal U}\right) \\
\Delta E_{{\cal D}_L, 3\times 3}^{(1)} & = & \left(\begin{array}{cc}
V_{Jq}^{{\cal D}\,\dagger} V_{Jq}^{\cal D} &
V_{Jq}^{{\cal D}\,\dagger} V_{JQ}^{\cal D} \\
V_{JQ}^{{\cal D}\,\dagger} V_{Jq}^{\cal D} \rule{0cm}{.5cm} &
V_{JQ}^{{\cal D}\,\dagger} V_{JQ}^{\cal D}
\end{array}\right) \ = \ \openone_{3\times 3} \,
- \, V_0^{{\cal D}\,\dagger}\, V_0^{\cal D} \\
\Delta E_{{\cal U}_L, 3\times 3}^{(2)} & = & - \ 2\,C_W^2
\left(\begin{array}{cc}
V_{Jq}^{{\cal U}\,\dagger} V_{Jq}^{\cal U} +
V_{Qq}^{{\cal U}\,\dagger} V_{Qq}^{\cal U} &
V_{Jq}^{{\cal U}\,\dagger} V_{JQ}^{\cal U} +
V_{Qq}^{{\cal U}\,\dagger} V_{QQ}^{\cal U} \\
V_{JQ}^{{\cal U}\,\dagger} V_{Jq}^{\cal U} +
V_{QQ}^{{\cal U}\,\ast} V_{Qq}^{\cal U} \rule{0cm}{.5cm} &
V_{JQ}^{{\cal U}\,\dagger} V_{JQ}^{\cal U} +
V_{QQ}^{{\cal U}\,\ast} V_{QQ}^{\cal U}
\end{array}\right) \\
& & + \ (C_W^2 \mp 2C_W^2 \pm S_W^2)
\left(\openone_{3\times 3} \, -
\, V_0^{{\cal U}\,\dagger}\, V_0^{\cal U}\right)
\\
\Delta E_{{\cal D}_L, 3\times 3}^{(2)} & = & - \ 2\,C_W^2
\left(\begin{array}{cc}
V_{Jq}^{{\cal D}\,\dagger} V_{Jq}^{\cal D} +
V_{Qq}^{{\cal D}\,\dagger} V_{Qq}^{\cal D} &
V_{Jq}^{{\cal D}\,\dagger} V_{JQ}^{\cal D} +
V_{Qq}^{{\cal D}\,\dagger} V_{QQ}^{\cal D} \\
V_{JQ}^{{\cal D}\,\dagger} V_{Jq}^{\cal D} +
V_{QQ}^{{\cal D}\,\ast} V_{Qq}^{\cal D} \rule{0cm}{.5cm} &
V_{JQ}^{{\cal D}\,\dagger} V_{JQ}^{\cal D} +
V_{QQ}^{{\cal D}\,\ast} V_{QQ}^{\cal D}
\end{array}\right) \\
& & + \ (C_W^2 \pm 2C_W^2 \mp S_W^2)
\left(\openone_{3\times 3} \, -
\, V_0^{{\cal D}\,\dagger}\, V_0^{\cal D}\right) \ ,
\end{eqnarray}
where double signs correspond to $\beta=\pm 1/\sqrt{3}$. We can proceed in
a similar way for the FCNC driven by the $K$ vector bosons. In order to
have a uniform notation we keep the definition in Eq.~(\ref{efl}), with
$E_{f_L}^{(i,{\rm diag})} = 0$ for $i=3,4$. Thus for $\beta=1/\sqrt{3}$
we get
\begin{eqnarray}
\Delta E_{{\cal U}_L, 3\times 3}^{(3)} & = &
\left(\begin{array}{cc}
V_{qq}^{{\cal U}\,\dagger} V_{Jq}^{\cal U} +
V_{Jq}^{{\cal U}\,\dagger} V_{qq}^{\cal U} &
V_{qq}^{{\cal U}\,\dagger} V_{JQ}^{\cal U} +
V_{Jq}^{{\cal U}\,\dagger} V_{qQ}^{\cal U} \\
V_{qQ}^{{\cal U}\,\dagger} V_{Jq}^{\cal U} +
V_{JQ}^{{\cal U}\,\dagger} V_{qq}^{\cal U} \rule{0cm}{.5cm} &
V_{qQ}^{{\cal U}\,\dagger} V_{JQ}^{\cal U} +
V_{JQ}^{{\cal U}\,\dagger} V_{qQ}^{\cal U}
\end{array}\right) \label{e3u} \\
\Delta E_{{\cal D}_L, 3\times 3}^{(3)} & = &
\left(\begin{array}{cc}
V_{Qq}^{{\cal D}\,\dagger} V_{Jq}^{\cal D} +
V_{Jq}^{{\cal D}\,\dagger} V_{Qq}^{\cal D} &
V_{Qq}^{{\cal D}\,\dagger} V_{JQ}^{\cal D} +
V_{Jq}^{{\cal D}\,\dagger} V_{QQ}^{\cal D} \\
V_{QQ}^{{\cal D}\,\ast} V_{Jq}^{\cal D} +
V_{JQ}^{{\cal D}\,\dagger} V_{Qq}^{\cal D} \rule{0cm}{.5cm} &
V_{QQ}^{{\cal D}\,\ast} V_{JQ}^{\cal D} +
V_{JQ}^{{\cal D}\,\dagger} V_{QQ}^{\cal D}
\end{array}\right) \\
\Delta E_{{\cal U}_L, 3\times 3}^{(4)} & = &
\left(\begin{array}{cc}
V_{qq}^{{\cal U}\,\dagger} V_{Jq}^{\cal U} -
V_{Jq}^{{\cal U}\,\dagger} V_{qq}^{\cal U} &
V_{qq}^{{\cal U}\,\dagger} V_{JQ}^{\cal U} -
V_{Jq}^{{\cal U}\,\dagger} V_{qQ}^{\cal U} \\
V_{qQ}^{{\cal U}\,\dagger} V_{Jq}^{\cal U} -
V_{JQ}^{{\cal U}\,\dagger} V_{qq}^{\cal U} \rule{0cm}{.5cm} &
V_{qQ}^{{\cal U}\,\dagger} V_{JQ}^{\cal U} -
V_{JQ}^{{\cal U}\,\dagger} V_{qQ}^{\cal U}
\end{array}\right) \\
\Delta E_{{\cal D}_L, 3\times 3}^{(4)} & = &
\left(\begin{array}{cc}
-V_{Qq}^{{\cal D}\,\dagger} V_{Jq}^{\cal D} +
V_{Jq}^{{\cal D}\,\dagger} V_{Qq}^{\cal D} &
-V_{Qq}^{{\cal D}\,\dagger} V_{JQ}^{\cal D} +
V_{Jq}^{{\cal D}\,\dagger} V_{QQ}^{\cal D} \\
-V_{QQ}^{{\cal D}\,\ast} V_{Jq}^{\cal D} +
V_{JQ}^{{\cal D}\,\dagger} V_{Qq}^{\cal D} \rule{0cm}{.5cm} &
-V_{QQ}^{{\cal D}\,\ast} V_{JQ}^{\cal D} +
V_{JQ}^{{\cal D}\,\dagger} V_{QQ}^{\cal D}
\end{array}\right) \ , \label{e4d}
\end{eqnarray}
while for $\beta=-1/\sqrt{3}$ one obtains similar expressions, changing
$\Delta E_{{\cal U}_L, 3\times 3}^{(3,4)} \to -\Delta E_{{\cal D}_L, 3\times
3}^{(3,4)}$ and $\Delta E_{{\cal D}_L, 3\times 3}^{(3,4)} \to \Delta
E_{{\cal U}_L, 3\times 3}^{(3,4)}$.

Our results can be compared with those obtained in
Ref.~\cite{Langacker:2000ju}, where the authors study FCNC driven by a $Z'$
boson associated with an additional abelian gauge symmetry. The model in
that work does not consider the presence of exotic fermions, nor that of a
gauge boson analogous to the $K$, therefore in general the structure of FCNC
is different from that of the 331 models. However, it is easily seen that in
the limit where both the mixing between ordinary and exotic quarks and the
mixing between $K$ and $Z$, $Z'$ gauge bosons are neglected, our expressions
reduce to a particular case of those quoted in Ref.~\cite{Langacker:2000ju}.
Namely, the matrices $B^{\psi_{L,R}}$ of Ref.~\cite{Langacker:2000ju},
responsible for flavor-changing effects, are nothing but our flavor matrices
$E_{f_{L,R}}^{(2)}$, and the corresponding coupling $g_2$ is that given by
Eq.~(\ref{gs}). Therefore, in the above mentioned limit, the
phenomenological analysis carried out in Ref.~\cite{Langacker:2000ju} is
directly applicable to 331 models.

\hfill

As stated, the elements of the matrices $V_L^f$ can be obtained from the
Yukawa coupling constants and the VEVs of the neutral scalar fields. Since
these parameters are not known, in practice the only conditions to be
satisfied are the unitarity of $V_L^{{\cal U},{\cal D}}$ and the
experimental constraints on $V_{CKM} = V_0^{{\cal U}\,\dagger}V_0^{\cal
D}$ (these constraints have to be taken with care, since $V_{CKM}$ is in
general not unitary). However, most theoretical models of fermion mass
matrices lead to hierarchies in quark mixing angles, which are related to
the corresponding hierarchies in quark masses. This is clearly inspired in
the experimental measurements of $V_{CKM}$, which is surprisingly close to
the identity. According to this picture not only the product $V_0^{{\cal
U}\,\dagger} V_0^{\cal D}$ would be close to the identity but also
$V_0^{{\cal U}\,\dagger}$ and $V_0^{\cal D}$ separately.

In order to analyze the phenomenological consequences of these mass matrix
structures in the context of the 331 theories, let us study the dominant
contributions to tree level FCNC. For definiteness we will consider the
case in which off-diagonal elements of mixing matrices satisfy family
hierarchies given by
\begin{equation}
V^f_{\,ij} \ \sim \ \left(\frac{m_{f_i}}{m_{f_j}}\right)^{1/2} \ .
\label{assum}
\end{equation}
These hierarchies in $V^f_{ij}$ are obtained in various models for quark
mass matrices, which have been widely studied in the
literature~\cite{Fritzsch,Cheng:1987rs}. Notice that the assumption of
this ansatz for the mixing submatrices $V_0^{{\cal U},{\cal D}}$ in the
quark basis defined in Eq.~(\ref{content}) corresponds to a picture of
approximate flavor symmetries in which the third family of ordinary quarks
is associated with the states $u_3$ and $d_3$. This is also suggested by
the notation in Eq.~(\ref{submat}), where we have labelled with $Q$ and
$q$ quarks belonging to the $Q_{3L}$ and $Q_{1,2\,L}$ multiplets
respectively. In fact, the third family could be associated to any
combination of $Q_{1L}$, $Q_{2L}$ and $Q_{3L}$. However, once approximate
flavor symmetries are considered, it is natural to assume that the family
that distinguishes from the other two [in the sense that the states
transform according to a different $SU(3)$ representation] is the third
(heavier) family of ordinary quarks.

It is interesting to notice that, within this framework, one can distinguish
between FCNC in the up and down quark sectors. For the up quark sector,
given the small ratios $m_u/m_t$ and $m_c/m_t$, Eq.~(\ref{assum}) implies
values of order $\lesssim 10^{-1}$ for the elements of $V^{\cal U}_{qQ}$ and
$V^{\cal U}_{Qq}$, which mix quarks $u,c$ with quark $t$. Here it is
interesting to consider the case in which the $SU(3)_L\otimes U(1)_X$
breaking scale is of the order of 1 TeV. In this case one expects exotic
quarks $T$ to have masses of the TeV order, therefore the elements of
$V^{\cal U}_{QJ}$ and $V^{\cal U}_{JQ}$ can be relatively large. In this
way, one has
\begin{eqnarray}
\Delta E_{{\cal U}_L, 3\times 3}^{(1)} & \simeq &
- \left( \begin{array}{cc}
V_{Jq}^{{\cal U}\,\dagger} V_{Jq}^{\cal U} &
V_{Jq}^{{\cal U}\,\dagger} V_{JQ}^{\cal U} \\
V_{JQ}^{{\cal U}\,\dagger} V_{Jq}^{\cal U} \rule{0cm}{.5cm} &
1 - |V^{\cal U}_{QQ}|^2
\end{array}\right) \label{ape1u} \\
\Delta E_{{\cal U}_L, 3\times 3}^{(2)} & \simeq & - \ 2\,C_W^2
\left(\begin{array}{cc}
V_{Qq}^{{\cal U}\,\dagger} V_{Qq}^{\cal U} &
V_{Qq}^{{\cal U}\,\dagger} V_{QQ}^{\cal U} \\
V_{QQ}^{{\cal U}\,\ast} V_{Qq}^{\cal U} \rule{0cm}{.5cm} &
1
\end{array}\right) \ + \ (C_W^2 \mp 2C_W^2 \pm S_W^2)
\left( \begin{array}{cc}
0_{2\times 2} & 0_{2\times 1}  \\
0_{1\times 2} \rule{0cm}{.5cm} &
1 - |V^{\cal U}_{QQ}|^2
\end{array}\right) \ ,
\label{ape2u}
\end{eqnarray}
where we have used the relation $V^{{\cal U}\,\dagger}_{JQ} V^{\cal U}_{JQ}
+ |V^{\cal U}_{QQ}|^2 = 1 - V^{{\cal U}\,\dagger}_{qQ} V^{\cal U}_{qQ}
\simeq 1$. Eqs.~(\ref{ape1u}) and (\ref{ape2u}) show that $Z$ and
$Z'$-mediated FCNC in the up quark sector are expected to be suppressed by
the small values of $V^{\cal U}_{Jq}$ and $V^{\cal U}_{Qq}$, respectively.

On the other hand, in the down quark sector it is natural to expect $m_b\ll
m_B$, therefore ordinary quarks decouple, and both $qJ$ and $QJ$ mixing will
be suppressed. The submatrix $V_0^{\cal D}$ can be taken in this case as
approximately unitary. However, one can expect a significant mixing between
$b$ and $s$ quarks ($\sqrt{(m_s/m_b)}\simeq 0.15$), leading to nonnegligible
values of the matrix elements of $V_{qQ}$ and $V_{Qq}$. One has then
\begin{eqnarray}
\Delta E_{{\cal D}_L, 3\times 3}^{(1)} & = & \openone_{3\times 3} \,
- \, V_0^{{\cal D}\,\dagger}\, V_0^{\cal D} \\
\Delta E_{{\cal D}_L, 3\times 3}^{(2)} & \simeq & - \ 2\,C_W^2
\left(\begin{array}{cc}
V_{Qq}^{{\cal D}\,\dagger} V_{Qq}^{\cal D} &
V_{Qq}^{{\cal D}\,\dagger} V_{QQ}^{\cal D} \\
V_{QQ}^{{\cal D}\,\ast} V_{Qq}^{\cal D} \rule{0cm}{.5cm} &
V_{QQ}^{{\cal D}\,\ast} V_{QQ}^{\cal D}
\end{array} \right) \ .
\label{edl}
\end{eqnarray}
Since, as stated, $V_0^{\cal D}$ is approximately unitary, it is seen that
$Z$-mediated FCNC should be suppressed. On the other hand, the
off-diagonal matrix elements in (\ref{edl}) could be significant enough so
that the size of $Z'$-mediated FCNC in the down sector could be comparable
with (one-loop) SM contributions.

Notice that the effect of $QJ$ mixing in the up quark sector should be taken
into account when considering the unitarity of the $V_{CKM}$ matrix. Indeed,
neglecting $qQ$, $qJ$ mixing in the up quark sector, as well as $qJ$, $QJ$
mixing in the down quark sector, one finds
\begin{equation}
V_{CKM}^\dagger\; V_{CKM} \ = \
\left(V_0^{{\cal U}\,\dagger}\; V_0^{\cal D}\right)^\dagger\;
V_0^{{\cal U}\,\dagger}\; V_0^{\cal D} \ \simeq \
\left( \begin{array}{cc}
\openone_{2\times 2} &   \\
 \rule{0cm}{.5cm} &
1 - V^{{\cal U}\,\dagger}_{JQ} V^{\cal U}_{JQ}
\end{array}\right) \ .
\label{vckmunit}
\end{equation}
This deviation from unitarity could be a signature for the presence of new
physics. In particular, notice that Eq.~(\ref{vckmunit}) implies
$|V_{ub}|^2+|V_{cb}|^2+|V_{tb}|^2 < 1$, which could be tested from direct
measurements of the corresponding matrix elements. This test would require
the analysis of the decay $t\to Wb$, which could be achieved in forthcoming
experiments such as the LHC.

Finally, from Eqs.~(\ref{e3u}-\ref{e4d}) one can determine the dominant
contributions to tree level FCNC mediated by $K$ vector bosons. For $\beta =
1/\sqrt{3}$ one has
\begin{equation}
\Delta E_{{\cal U}_L, 3\times 3}^{(3)}
\ \simeq \
\left(\begin{array}{cc}
& V_{JQ}^{\cal U} \\
V_{JQ}^{{\cal U}\,\dagger} \rule{0cm}{.5cm} &
\end{array}\right) \ , \qquad
\Delta E_{{\cal U}_L, 3\times 3}^{(4)} \ \simeq \
\left(\begin{array}{cc}
& V_{JQ}^{\cal U} \\
- V_{JQ}^{{\cal U}\,\dagger} \rule{0cm}{.5cm} &
\end{array}\right) \ ,
\end{equation}
while for $\beta = - 1/\sqrt{3}$ we get
\begin{equation}
\Delta E_{{\cal U}_L, 3\times 3}^{(3)}
\ \simeq \
\left(\begin{array}{cc}
0_{2\times 2} & \\
& 2\,{\rm Re}(V_{QQ}^{{\cal U}\,\ast}V_{JQ}^{\cal U})
\rule{0cm}{.5cm}
\end{array}\right) \ , \qquad
\Delta E_{{\cal U}_L, 3\times 3}^{(4)} \ \simeq \
\left(\begin{array}{cc}
0_{2\times 2} & \\
& -2i\,{\rm Im}(V_{QQ}^{{\cal U}\,\ast}V_{JQ}^{\cal U})
\rule{0cm}{.5cm}
\end{array}\right) \ .
\end{equation}
For both values of $\beta$ the elements of $\Delta E_{{\cal D}_L, 3\times
3}^{(3,4)}$ will be strongly suppressed, according to the above
assumptions on Yukawa coupling hierarchies. From the above equations it is
seen that only in the up quark sector and for $\beta = 1/\sqrt{3}$ one
would expect sizable $K$-mediated FCNC. Though the disentanglement of the
$K$ contribution from the $Z$ and $Z'$ ones could be challenging, it is
noticeable that it would offer the opportunity to distinguish between
models with different choices of $\beta$ without the direct observation of
exotic quarks or leptons.

\section{Neutral meson mass differences}

As an application of the formalism presented above, let us analyze the
effective lagrangian for $\Delta F=2$ four-quark couplings ($F$ here for
flavor) arising from the above described FCNC. This effective interaction
is particularly important since $\Delta F=2$ observables, such as neutral
meson mass differences in the $K$, $B$ and $D$ sectors, provide stringent
bounds for FCNC~\cite{GomezDumm:1994tz,Promberger:2007py,Cabarcas:2007my}.

If we denote the flavor eigenstates of a neutral meson system by $P^0$ and
$\overline{P}^0$, and the corresponding mass eigenstates by $P_1$ and $P_2$,
the mass difference $\Delta m_P = m_{P_1}-m_{P_2}$ will be given by
\begin{equation}
\Delta m_P \ = \ \frac{{\rm Re}\,\langle P^0\,|{\cal H}_{\rm
eff}^{(\Delta F=2)}|\,\overline{P}^0\rangle}{m_P} \ .
\end{equation}
With the notation introduced in previous sections, the effective Lagrangian
in 331 models can be written as
\begin{eqnarray}
{\cal L}_{\rm eff}^{(\Delta F=2)} &  = & - \frac{4\,G_F}{\sqrt2} \sum_{f = {\cal
U},{\cal D}} \ \sum_{a\neq b} C^{(f)}_{ab}\; \bar{f}_a\, \gamma^\mu\,
P_L\, f_b \ \bar{f}_a\, \gamma^\mu\, P_L\, f_b \ ,
\label{leff}
\end{eqnarray}
where $a,b=1,2,3$, and
\begin{equation}
C^{(f)}_{ab} \ = \ \sum_{j,k,l=1}^3\; \left[ \frac{g_j\, g_l}{g^2} \;
(\Delta E_{f_L}^{(j)})_{ab}\;
(\Delta E_{f_L}^{(l)})_{ab}\;
R_{jk}\,R_{lk}\; \frac{m_W^2}{m_{Z_k}^2} \right]
\ - \ \frac{m_W^2}{4\, m_{Z_k}^2}\;
[(\Delta E_{f_L}^{(4)})_{ab}]^2 \ .
\label{cab}
\end{equation}

In this way, the contribution from tree level FCNC to the $P_1-P_2$ mass
difference is given by
\begin{equation}
\label{dmas1}
\Delta m_P \ = \ \frac{4\sqrt{2}}{3}\,G_F\, m_P\, F_P^2\, B_P\,
{\rm Re} [C_{ab}^{(f)}] \ ,
\end{equation}
where $f$ is ${\cal U}$ or ${\cal D}$ depending on the flavor content of
$P$, and $B_P$ parameterizes the theoretical uncertainty in the evaluation
of the hadronic matrix elements. Within the vacuum insertion approximation
one has $B_P=1$.

Now it is interesting to consider the effective Lagrangian in
Eqs.~(\ref{leff}) and (\ref{cab}) to distinguish between different 331
scenarios:

\begin{itemize}

\item Although we have concentrated in 331 models with $\beta = \pm
1/\sqrt{3}$, it is also interesting to analyze the models with $\beta =
\pm\sqrt{3}$~\cite{Pisano:1991ee}. The latter (which, actually, were
proposed firstly in the literature) include exotic quarks that carry
exotic electric charges, namely $5/3$ and $-4/3$, therefore they do not
mix with ordinary quarks. Moreover, there are no neutral $K$ vector
bosons. Since $V_L^{\cal U,D}$ reduce to $V_0^{\cal U,D}$, which are now
unitary, there is no FCNC mediated by the $Z$ boson. Eq.~(\ref{cab})
reduces in this case to
\begin{equation}
C^{(f)}_{ab} \ = \ \frac{m_W^2}{4\,C_W^2}\;
\left(\frac{g_2}{g_1}\right)^2
[(\Delta E_{f_L}^{(2)})_{ab}]^2\;
\left( \frac{\sin^2\theta}{m_{Z_1}^2}
\; +\; \frac{\cos^2\theta}{m_{Z_2}^2} \right) \ ,
\label{rtres}
\end{equation}
where
\begin{equation}
\Delta E_{f_L}^{(2)} \ = \
- \ 2\,C_W^2
\left(\begin{array}{cc}
V_{Qq}^{f\,\dagger} V_{Qq}^f &
V_{Qq}^{f\,\dagger} V_{QQ}^f \\
V_{QQ}^{f\,\ast} V_{Qq}^f \rule{0cm}{.5cm} &
V_{QQ}^{f\,\ast} V_{QQ}^f
\end{array}\right) \ ,
\end{equation}
and $\theta$ is the mixing angle in the $2\times 2$ matrix $R$ that mixes
$Z$ and $Z'$ states ($R_{12}=-R_{21}=\sin\theta$). As it is shown e.g.\ in
Ref.~\cite{Liu:1993fwa}, this angle is suppressed by the VEV scale ratio
$(v/V)^2$. Bounds for 331 model parameters have been determined in this
context in Ref.~\cite{Promberger:2007py} from various FCNC observables. It
is also important to notice that the expression in Eq.~(\ref{rtres}) can be
obtained from the analysis in Ref.~\cite{Langacker:2000ju}, applied to the
case of the 331 model as discussed in the previous section.

In addition, it is worth to notice that
\begin{equation}
\left(\frac{g_2}{g_1}\right)^2
\ = \
\frac{1}{3\,(C_W^2-\beta^2\,S_W^2)} \ ,
\end{equation}
hence for $\beta =\pm\sqrt{3}$ the ratio is divergent for $S_W^2=1/4$, where
one finds a Landau-like pole for the coupling $g'$. If one requires the
interaction to be perturbative, the presence of this pole imposes in general
an upper bound of a few TeV for the symmetry breaking scale
$V$~\cite{Dias:2004dc} (this bound can be somewhat evaded, however, by
enlarging the particle content of the model~\cite{Dias:2004wk}). Models with
$\beta= \pm 1/\sqrt{3}$ are free from this constraint. A brief comparative
analysis of FCNC in the $B$ meson sector for different choices of $\beta$
can be found in Ref.~\cite{Rodriguez:2004mw}.

\item Let us consider 331 models with $\beta = \pm 1/\sqrt{3}$, assuming
a quark mixing hierarchy such as that in Eq.~(\ref{assum}). As stated
above, in this case $Z$- and $K$-mediated FCNC in the down quark sector
are negligible. One has then
\begin{equation}
C^{({\cal D})}_{ab} \ \simeq \ \frac{m_W^2}{4C_W^2(3C_W^2-S_W^2)}\;
[(\Delta E_{{\cal D}_L}^{(2)})_{ab}]^2\;
\sum_{k=1}^3 \frac{R_{2k}^2}{m_{Z_k}^2} \ .
\end{equation}
Moreover, in general one expects the mixing between $Z$ and $Z'$ states,
i.e.\ $R_{21}$, to be of order $(v/V)^2$, therefore the dominant effect is
due to the contribution of the (heavy) $Z_2$ intermediate state. As shown in
Ref.~\cite{Rodriguez:2004mw}, the stringent bounds for the $Z_2$ mass are
then provided by the measured values of $\Delta m_{B_d}$ and $\Delta
m_{B_s}$. From the above expressions, taking into account the result
in Eq.~(\ref{edl}) one gets
\begin{equation}
\Delta m_{B_q}^{(Z_2)} \ \simeq \ \frac{4\sqrt{2}}{3}\,
\frac{C_W^2}{3\,C_W^2-S_W^2} \; G_F\, m_{B_q}\, F_{B_q}^2\,
B_{B_q}\,\frac{m_q}{m_b}\,\left(\frac{m_W}{m_{Z_2}}\right)^2
 \ , \qquad q=d,s \ .
\end{equation}
Present experimental results for these observables lead to $m_{Z_2}\gtrsim
7$ TeV. Clearly, this bound can be relaxed either if one considers
interferences with SM one-loop contributions or other sources of FCNC, or
if one relaxes the hierarchy relation in the quark mixing angles. Within
the economical 331 model, a general analysis of the constraints for masses
and mixing angles from $\Delta F=2$ observables has been carried out in
Ref.~\cite{Cabarcas:2007my}.

On the other hand, in the up quark sector, for $\beta=1/\sqrt{3}$ the
largest FCNC couplings are those driven by the neutral $K$ state,
subsequently followed by those driven by the $Z'$ and the $Z$. Thus,
depending on the gauge boson masses and mixing angles, the contributions
of $Z_1$, $Z_2$ and $Z_3$ could be competitive and all terms in
(\ref{cab}) should be taken into account. For $\beta = -1/\sqrt{3}$, given
the large $K$ mass, the contribution of the $K$ state is clearly
suppressed in comparison with that of the $Z$, therefore $\Delta E_{{\cal
U}_L}^{(3,4)}$ can be neglected, and only $Z$ and $Z'$ contributions
compete. The presence of tree level FCNC in the up quark sector leads to
nonstandard contributions to the $\Delta m_D$ mass difference, which has
been recently measured~\cite{Aubert:2007wf} attracting significant
theoretical interest. Considering the quark mixing hierarchy in
Eq.~(\ref{assum}), and neglecting the mixing between neutral gauge boson
states, one can obtain a definite prediction for the order of magnitude of
the new contributions. From Eqs.~(\ref{ape1u}) and (\ref{ape2u}) we get
\begin{eqnarray}
\Delta m_{D}^{(Z_1)} & \simeq & \frac{\sqrt{2}}{3}\; G_F\, m_{D}\,
F_{D}^2\, B_{D} \; \frac{m_u\, m_c}{m_T^2} \nonumber \\
\Delta m_{D}^{(Z_2)} & \simeq & \frac{\sqrt{2}}{3}\; G_F\, m_{D}\,
F_{D}^2\, B_{D} \;
\frac{4\,C_W^2}{3\,C_W^2-S_W^2}\;\frac{m_u\, m_c}{m_t^2}
\,\left(\frac{m_W}{m_{Z_2}}\right)^2
\end{eqnarray}
($K$-mediated FCNC are negligible for this observable). If we consider the
upper bound of 7 TeV for $Z_2$ suggested by $B_d$ and $B_s$ mixing, the
contribution $\Delta m_{D}^{Z_2}$ from the above expression is found to be
about two orders of magnitude below the experimental value $\Delta
m_D^{(\rm exp)}=(2.4\pm 0.7)\times 10^{10}\hbar \,s^{-1}$. However, the
contribution mediated by the ordinary $Z_1$ boson could be significantly
larger if the mass of the exotic $T$ quark is in the TeV range. Taking
$B_D=0.82$~\cite{Gupta:1996yt} and $F_D=0.207$ GeV one has
\begin{equation}
\Delta m_{D}^{(Z_1)} \ \sim \ 2\times 10^9\; \hbar\,s^{-1}\,
\left(\frac{\rm 1\ TeV}{m_T}\right)^2\ ,
\end{equation}
thus it is seen that an observable effect could be achieved if the
experimental errors are reduced. It is clear that the above result has to be
taken just as an indicative order of magnitude. However, this rough
calculation illustrates the possible size of the effects one could expect in
331 models with $\beta=\pm 1/\sqrt{3}$ within the ansatz of quark mixing
hierarchy given by Eq.~(\ref{assum}).

\item If spontaneous CP violation is allowed, Eq.~(\ref{cab}) should still
be generalized to allow for the mixing between $Z_{1,2,3}$ and $Z_4$. This
can be implemented in Eq.~(\ref{cab}) by removing the last term, enlarging
$R$ to a $4\times 4$ orthogonal matrix and extending the sum up to four. The
entries $R_{i4}$, $R_{4i}$ with $i=1,2,3$ will be responsible for CP
violation. These elements will arise from the imaginary parts of scalar
VEVs, which are in general nonzero if CP is spontaneously broken together
with the gauge symmetry~\cite{Epele:1995vv}. However, notice that if the
scalar sector is minimal (i.e.\ in the case of ``economical'' 331 models)
VEV phases can be removed through a global gauge transformation. In this
case $Z_4$ decouples, becoming an exact mass eigenstate.

\end{itemize}

\section{Summary}

In summary, we present here a general analysis of tree level flavor
changing neutral currents in the context of 331 models. We quote the
corresponding expressions for arbitrary quark and gauge boson mixing
matrices, and then analyze the approximate results obtained for a definite
texture of the quark mass matrices. Differences between various 331 models
are pointed out.

Our analysis generalizes previous results, including FCNC contributions
that are usually neglected in the literature, such as those arising from
the mixing between ordinary and exotic quarks and the mixing between
neutral gauge bosons. Even if these contributions are in most cases
expected to be suppressed, it is seen that they can be significant for
some observables, such as e.g.\ the mass difference between neutral
$D$ meson states.

\section{Acknowledgements}
This work has been supported in part by Banco de la Rep\'ublica (Colombia)
and by CONICET and ANPCyT (Argentina), under grants PIP 6009, PIP
112-200801-02495, PICT04-03-25374 and PICT07-03-00818.


\begin{thebibliography}{50}

\bibitem{Pisano:1991ee}
  F.~Pisano and V.~Pleitez,
  Phys.\ Rev.\  D {\bf 46}, 410 (1992);
  P.~H.~Frampton,
  Phys.\ Rev.\ Lett.\  {\bf 69}, 2889 (1992);
  R.~Foot, O.~F.~Hernandez, F.~Pisano and V.~Pleitez,
  Phys.\ Rev.\  D {\bf 47}, 4158 (1993).

\bibitem{Ng:1992st}
  D.~Ng,
  Phys.\ Rev.\  D {\bf 49}, 4805 (1994).

\bibitem{kiyan:2002qw}
  T.~kiyan, T.~Maekawa and S.~Yokoi,
  Mod.\ Phys.\ Lett.\  A {\bf 17}, 1813 (2002).

\bibitem{Diaz:2004fs}
  R.~A.~Diaz, R.~Martinez and F.~Ochoa,
  Phys.\ Rev.\  D {\bf 72}, 035018 (2005).

\bibitem{Montero:2000rh}
  J.~C.~Montero, C.~A.~de S. Pires and V.~Pleitez,
  Phys.\ Lett.\  B {\bf 502}, 167 (2001);
  Phys.\ Rev.\  D {\bf 65}, 095001 (2002).

\bibitem{Fregolente:2002nx}
  D.~Fregolente and M.~D.~Tonasse,
  Phys.\ Lett.\  B {\bf 555}, 7 (2003);
  L.~N.~Hoang and N.~Q.~Lan,
  Europhys.\ Lett.\  {\bf 64}, 571 (2003);
  S.~Filippi, W.~A.~Ponce and L.~A.~Sanchez,
  Europhys.\ Lett.\  {\bf 73}, 142 (2006);
  C.~A.~de S.Pires and P.~S.~Rodrigues da Silva,
  JCAP {\bf 0712}, 012 (2007).

\bibitem{Diaz:2005bw}
  R.~A.~Diaz, D.~Gallego and R.~Martinez,
  Int.\ J.\ Mod.\ Phys.\  A {\bf 22}, 1849 (2007).

\bibitem{Liu:1993gy}
  J.~T.~Liu and D.~Ng,
  Phys.\ Rev.\  D {\bf 50}, 548 (1994);
  J.~T.~Liu,
  Phys.\ Rev.\  D {\bf 50}, 542 (1994).

\bibitem{GomezDumm:1994tz}
  D.~Gomez Dumm, F.~Pisano and V.~Pleitez,
  Mod.\ Phys.\ Lett.\  A {\bf 9}, 1609 (1994).

\bibitem{Rodriguez:2004mw}
  J.~A.~Rodriguez and M.~Sher,
  Phys.\ Rev.\  D {\bf 70}, 117702 (2004).

\bibitem{Promberger:2007py}
  C.~Promberger, S.~Schatt and F.~Schwab,
  Phys.\ Rev.\  D {\bf 75}, 115007 (2007).

\bibitem{Cabarcas:2007my}
  J.~M.~Cabarcas, D.~Gomez Dumm and R.~Martinez,
  Phys.\ Rev.\  D {\bf 77}, 036002 (2008).

\bibitem{Promberger:2008xg}
  C.~Promberger, S.~Schatt, F.~Schwab and S.~Uhlig,
  Phys.\ Rev.\  D {\bf 77}, 115022 (2008).

\bibitem{Martinez:2008jj}
  R.~Martinez and F.~Ochoa,
  Phys.\ Rev.\  D {\bf 77}, 065012 (2008).

\bibitem{Cabarcas:2008ys}
  J.~M.~Cabarcas, D.~Gomez Dumm and R.~Martinez,
  Eur.\ Phys.\ J.\  C {\bf 58}, 569 (2008).

\bibitem{Dittmar:2003ir}
  M.~Dittmar, A.~S.~Nicollerat and A.~Djouadi,
  Phys.\ Lett.\  B {\bf 583}, 111 (2004).

\bibitem{Langacker:2000ju}
  P.~Langacker and M.~Plumacher,
  Phys.\ Rev.\  D {\bf 62}, 013006 (2000).

\bibitem{Nardi:1992nq}
  E.~Nardi,
  Phys.\ Rev.\  D {\bf 48}, 1240 (1993).

\bibitem{Perez:2004jc}
  M.~A.~Perez, G.~Tavares-Velasco and J.~J.~Toscano,
  Phys.\ Rev.\  D {\bf 69}, 115004 (2004).

\bibitem{Foot:1994ym}
  R.~Foot, L.~N.~Hoang and T.~A.~Tran,
  Phys.\ Rev.\  D {\bf 50}, 34 (1994);
  L.~N.~Hoang,
  Phys.\ Rev.\  D {\bf 53}, 437 (1996);
  L.~N.~Hoang,
  Phys.\ Rev.\  D {\bf 54}, 4691 (1996).

\bibitem{Ozer:1995xi}
  M.~Ozer,
  Phys.\ Rev.\  D {\bf 54}, 1143 (1996).

\bibitem{Foot:1992rh}
  R.~Foot, O.~F.~Hernandez, F.~Pisano and V.~Pleitez,
  Phys.\ Rev.\  D {\bf 47}, 4158 (1993).

\bibitem{Montero:2001tq}
  J.~C.~Montero, C.~A.~de S. Pires and V.~Pleitez,
  Phys.\ Rev.\  D {\bf 66}, 113003 (2002).

\bibitem{Nguyen:1998ui}
  T.~A.~Nguyen, N.~A.~Ky and L.~N.~Hoang,
  Int.\ J.\ Mod.\ Phys.\  A {\bf 15}, 283 (2000).

\bibitem{Tully:1998wa}
  M.~B.~Tully and G.~C.~Joshi,
  Int.\ J.\ Mod.\ Phys.\  A {\bf 18}, 1573 (2003).

\bibitem{Diaz:2003dk}
  R.~A.~Diaz, R.~Martinez and F.~Ochoa,
  Phys.\ Rev.\  D {\bf 69}, 095009 (2004).

\bibitem{Ponce:2002sg}
  W.~A.~Ponce, Y.~Giraldo and L.~A.~Sanchez,
  Phys.\ Rev.\  D {\bf 67}, 075001 (2003).

\bibitem{Dong:2006mg}
  P.~V.~Dong, L.~N.~Hoang, D.~T.~Nhung and D.~V.~Soa,
  Phys.\ Rev.\  D {\bf 73}, 035004 (2006).

\bibitem{Cotaescu:2008ya}
  I.~I.~Cotaescu and A.~Palcu,
  Mod.\ Phys.\ Lett.\  A {\bf 23}, 1011 (2008).

\bibitem{Fritzsch}
H. Fritzsch, Phys.\ Lett.\ B {\bf 73}, 317 (1978); Nucl.\ Phys.\ B {\bf
155}, 189 (1979).

\bibitem{Cheng:1987rs}
  T.~P.~Cheng and M.~Sher,
  Phys.\ Rev.\  D {\bf 35}, 3484 (1987).

\bibitem{Liu:1993fwa}
  J.~T.~Liu and D.~Ng,
  Z.\ Phys.\  C {\bf 62}, 693 (1994).

\bibitem{Dias:2004dc}
  A.~G.~Dias, R.~Martinez and V.~Pleitez,
  Eur.\ Phys.\ J.\  C {\bf 39}, 101 (2005).

\bibitem{Dias:2004wk}
  A.~G.~Dias,
  Phys.\ Rev.\  D {\bf 71}, 015009 (2005).

\bibitem{Aubert:2007wf}
  B.~Aubert {\it et al.}  [BABAR Collaboration],
  Phys.\ Rev.\ Lett.\  {\bf 98}, 211802 (2007);
  M.~Staric {\it et al.}  [Belle Collaboration],
  Phys.\ Rev.\ Lett.\  {\bf 98}, 211803 (2007).

\bibitem{Gupta:1996yt}
  R.~Gupta, T.~Bhattacharya and S.~R.~Sharpe,
  Phys.\ Rev.\  D {\bf 55}, 4036 (1997).

\bibitem{Epele:1995vv}
  L.~Epele, H.~Fanchiotti, C.~Garcia Canal and D.~Gomez Dumm,
  Phys.\ Lett.\  B {\bf 343}, 291 (1995);
  D.~Gomez Dumm,
  Int.\ J.\ Mod.\ Phys.\  A {\bf 11}, 887 (1996).

\end{thebibliography}
\end{document}